\documentclass[runningheads]{llncs}
\usepackage{makecell}
\usepackage{placeins}
\usepackage{multirow}
\usepackage{float}
\usepackage{graphicx}
\usepackage{graphics}
\usepackage{subcaption}
\usepackage[skins]{tcolorbox}
\usepackage[ruled,vlined]{algorithm2e}
\usepackage{amsmath}
\usepackage{appendix}
\usepackage{url}
\usepackage{color}
\usepackage{breqn}
\usepackage{pifont}
\usepackage{algpseudocode,caption}
\usepackage{float}
\usepackage{threeparttable}
\usepackage[inline]{enumitem}
\usepackage{booktabs}
\newcommand{\etal}{{\em et al.}\xspace}
\usepackage{xcolor}
\newcommand{\BfPara}[1]{{\vspace{0.3em}\noindent\bf#1.}\xspace}
\usepackage{xcolor}
\usepackage{colortbl}
\usepackage{tabularx}
\colorlet{lightgrey}{lightgray}
\usepackage{tikz}
\usepackage{pgf}
\usepackage{float}
\usepackage{authblk}

\usepackage{times}
\usepackage{tikz}

\usepackage{booktabs}
\usepackage{amsmath}
\usepackage{makecell}
\usepackage[colorlinks=true, allcolors=blue]{hyperref}
\usepackage{array}
\newcolumntype{L}[1]{>{\raggedright\let\newline\\\arraybackslash\hspace{0pt}}m{#1}}
\newcolumntype{C}[1]{>{\centering\let\newline\\\arraybackslash\hspace{0pt}}m{#1}}
\newcolumntype{R}[1]{>{\raggedleft\let\newline\\\arraybackslash\hspace{0pt}}m{#1}}

\begin{document}

\title{Understanding the Utilization of Cryptocurrency in the Metaverse and Security Implications}

\author[1]{Ayodeji Adeniran \and Mohammed Alkinoon \and David Mohaisen}
\institute{University of Central Florida, Orlando. USA}

\maketitle

\begin{abstract}
We present our results on analyzing and understanding the behavior and security of various metaverse platforms incorporating cryptocurrencies. We obtained the top metaverse coins with a capitalization of at least 25 million US dollars and the top metaverse domains for the coins, and augmented our data with name registration information (via whois), including the hosting DNS IP addresses, registrant location, registrar URL, DNS service provider, expiry date and check each metaverse website for information on fiat currency for cryptocurrency. The result from virustotal.com includes the communication files, passive DNS, referrer files, and malicious detections for each metaverse domain. Among other insights, we discovered various incidents of malicious detection associated with metaverse websites. Our analysis highlights indicators of (in)security, in the correlation sense, with the files and other attributes that are potentially responsible for the malicious activities.

\end{abstract}

\section{Introduction}\label{sec:introduction} 
Metaverse is a technology of the future with much anticipation and hype about its capabilities to alter the life of humans through online model values~\cite{buhalis2023metaverse}. Several companies are energetically working on building the metaverse, including technology giants like Facebook and Microsoft, among others.  The metaverse is still in its development phase, and the full realization of an interconnected virtual world is yet to be a reality. The metaverse holds the potential for various applications, such as entertainment, gaming, education, virtual commerce, virtual meetings, and more, and is expected to revolutionize how we socialize, work, learn, and interact with digital contents~\cite{akkus2022metaverse}. 

Although the metaverse is still developing, metaverse coins already amount to trillions of USD in value, and this trend is expected to persist as the technology reaches maturity~\cite{momtaz2022some}. However, as with any digital platform or online community~\cite{10.1007/978-3-031-26303-3_12}, the possibility of malicious activities occurring in the metaverse cannot be ignored.  As the metaverse concept evolves, it is essential to address potential security concerns, including detecting malicious activities within this virtual space. While the metaverse presents new opportunities for collaboration, interaction, and entertainment, it can also attract malicious actors who seek to exploit vulnerabilities or engage in harmful activities.  The intent and motivation for carrying out the malicious activity could be to steal vital information or assets that can be translated into money. Since the metaverse represents the digital world, which involves buying and selling with either cryptocurrency or fiat currency, malicious activities cannot be uncommon.
 
This paper focuses on understanding malicious activities in the metaverse represented by various platforms and domains. The attackers are sophisticated and experienced with reported attacks on other online platforms, e.g., cryptocurrencies and social media platforms. One of the ways the cyber attackers operate is by sending malicious files to the intended targets to corrupt the system and enable them to access it. The cyber-attacks can be malware~\cite{MohaisenAM15}, denial-of-service (DOS) attacks~\cite{WangMCC15}, phishing~\cite{ThomasM14}, or code injections~\cite{Mohaisen15}. Security analysis of the metaverse domains is the central focus of this paper, and we intend to analyze the files interacting with the domains to gain insight. We will discuss the possible security challenges and malicious activities in the metaverse.

\BfPara{Organization} In section~\ref{sec:related}, we present the related work, including the research gap. In section~\ref{sec:Problem Statement}, we introduce the problem statement, including the research questions. In section~\ref{sec:approach} we introduce our approach. In section~\ref{sec:results}, we discussed the results. We discuss various aspects of our studies in section~\ref{sec:discussion} and conclude our work in section~\ref{sec:conclusion}.

\section{Related Work}\label{sec:related}
Several papers explored the security of the metaverse. Di Pietro and Cresci~\cite{9750221} explored the security and privacy concerns surrounding the metaverse by focusing on the security risks that metaverse users may face and how it could affect their privacy. Zhao  \etal~\cite{ZHAO_2023} also conducted a study on security in the metaverse, discussing the common security issues and how they can impact the metaverse. Choi \etal~\cite{choi2022future} examined the future of the metaverse, tackled similar security issues as the previous ones, and discussed the technology and structural frameworks associated with the realization of solutions. 

Kurtunluoglu \etal~\cite{kurtunluouglu2022security} explored authentication in virtual reality and the metaverse, focusing on security and privacy concerns related to authentication methods. Aks \etal~\cite{aks2022review} also conducted a study on metaverse security, covering metaverse infrastructure, human interactions, and other interconnected virtual worlds aspects~\cite{jaber2022security}.

Tariq \etal~\cite{DBLP:journals/corr/abs-2303-14612} explored the security implications of deepfakes in the metaverse, the security challenges, authentication issues, and impersonation problems. Oosthoek \etal~\cite{DBLP:journals/tnsm/OosthoekD21} researched the security threats to cryptocurrencies, particularly to Bitcoin exchanges---Bitcoin is one of the major cryptocurrencies used in the metaverse. Zaghloul \etal~\cite{DBLP:journals/iotj/ZaghloulLM020} also examined the security and privacy issues with Bitcoin and blockchain relevant to the metaverse. Giechaskiel \etal~\cite{DBLP:journals/ieeesp/GiechaskielCR18} examined Bitcoin security challenges and their impact when there is a security breach or exposure.

Rosenberg \etal~\cite{DBLP:conf/uemcom/Rosnberg22} conducted a study on marketing in the metaverse and consumer protection. Rosenberg \etal~\cite{DBLP:conf/ficc/Rosenberg23} also studied marketing in the metaverse and the associated risks. Kshetri \etal~\cite{DBLP:journals/itpro/Kshetri23} studied the economics of the metaverse and its impact on the global economy. Other works that explored the security of cryptocurrencies in general include those in~\cite{SaadKM19,SaadCM21,SaadSurvey20}

\BfPara{The Research Gap} 
Our study is significantly different from other existing related studies. Our study examines the sources of security vulnerability in the metaverse and relates the findings to market capitalization. Unlike the prior work, our study conducts a thorough direct analysis of each metaverse token rather than focusing on general security concerns (e.g., human error, authentication issues, and other vulnerabilities). Our approach involves analyzing the top metaverse tokens, obtaining their domains and relevant information, and conducting a vulnerability scan to identify potential security issues that may lead to recommendations for this emerging application domain.

We note that our work is the first of its type in this space, as there is prior work that directly studied or measured the overlap between metaverse technologies and cryptocurrencies and how these cryptocurrencies are utilized within the metaverse.

\section{Problem Statement and Research Questions}\label{sec:Problem Statement} 
Both legal and illegal activities and transactions are expected in the metaverse. Metaverse is expected to become the digital center for gaming, entertainment, education, etc. Traffic to the metaverse will likely increase with millions of dollars in daily transactions. Security of assets, non-fungible tokens, cryptocurrency, and other technologies has become a challenge due to illegal activities associated with them in the metaverse.

To this end, this paper aims to tackle three crucial research questions related to identifying harmful behavior in the metaverse, particularly those associated with virtual tokens. Our analysis will be guided by these questions to ensure we provide accurate and self-contained answers. By scrutinizing various domains in the metaverse, we will obtain valuable insights that will aid our examination.

\begin{enumerate}
	\item {\bf RQ1: What are the prevalence of digital coins in the metaverse, and what are their associated threats?} We thoroughly scrutinize the correlation between the popularity and market capitalization of the metaverse and the plausible malicious threats. We analyzed the top forty metaverse coins with the highest market capitalization to accomplish this objective. %, what is their market capitalization, and how is that indicative of their potential threat?

	\item {\bf RQ2: How significant are metaverse domain artifacts such as communication and referring files in determining the maliciousness of such domains?} To effectively identify malicious incursions in Metaverse domains, conducting a thorough analysis of critical artifacts is imperative. This includes communication files, referrer files, and Passive DNS artifacts, which all directly impact Metaverse domains. Therefore, a comprehensive assessment of their contribution is essential.

	\item {\bf RQ3: Is there any correlation between fiat currency to cryptocurrency and vice versa, and the maliciousness of metaverse applications?} It is imperative to recognize the imminent threat posed by cyber attackers who aim to steal money and assets, especially in the metaverse, where cryptocurrency reigns supreme. Our investigation will determine whether domains incorporating fiat currency are more susceptible to malicious activities than those solely relying on cryptocurrency.

\end{enumerate}

\section{Technical Approach}\label{sec:approach}
This study explored the level of malicious activities in the top metaverse tokens. We analyzed 44 metaverse tokens with a market capitalization of at least 25 million USD. We hypothesize that cybercriminals are likelier to target tokens with a high market capitalization. To test this, we first divided the metaverse tokens into their respective domains and mapped them to their IP addresses. Then, we used the ``whois'' tool to gather information about the DNS service provider, registrar location and URL, hosting DNS IP addresses, and content delivery network (CDN). We manually inspected all the metaverse websites we studied for transactions from fiat to cryptocurrency.

We thoroughly scanned the metaverse domains and associated IP addresses using virustotal.com. During the scan, we gathered {\em passive DNS}, communication files, and referrer files and identified malicious detections. We then analyzed the communication and referrer files to detect any malicious activities and identified the file types to locate the source of the malicious activities. We then cross-referenced the metaverse domains with the malicious detections in the communication and referrer files to verify their presence. Additionally, we compared domains with fiat currency and cryptocurrency to domains with malicious activity. Lastly, we examined the metaverse tokens to identify patterns between the top and low tokens based on their market capitalizations.

\subsection{Dataset and Preprocessing}\label{sec:dataset}

\BfPara{Websites and Their Attributes} For this study, we collected data on metaverse coins, their corresponding domains, and their IP addresses. Our first step was to manually select metaverse coins with a market capitalization of at least 25 million USD and then map them to their respective domains. For the initial set of domains, we utilized \url{https://coinmarketcap.com}, a website that specializes in tracking coins, their market caps, and associated domains of application. To extract infrastructure information and address the first research question we posed in section~\ref{sec:Problem Statement}, we used domain query tools to extract information such as the IP addresses and CDN providers and {\em manually} checked each webpage for the presence of fiat currency. 

\BfPara{Security Data Attributes} We then scanned each metaverse domain and its associated IPs with \url{virustotal.com}. This scan provided information on Passive DNS, communication files, referrer files, and malicious detections. We further analyzed the communication files and referrer files to identify those with malicious detection and their types. The malicious detection was also categorized into different types with the number of occurrences for each type. Our primary focus was collecting data with malicious detection to explore the correlation between the different metaverse platforms, cryptocurrencies, artifacts, and associated malicious detection.

To gain a deeper understanding of file connections, especially those related to malicious activities, we thoroughly examined the interlinking between infected communication and referrer files and malware detections. Moreover, we meticulously tallied the frequency of each file type and its association with infected communication and referrer files. Our efforts to uncover malicious behavior were further amplified by our detailed analysis of every scan result and its correlation with malware detection in the scanned files and hosting metaverse platforms.

\subsection{Analysis Dimensions}
Our study explores the relationship between the metaverse domains and malicious activity and detection. We aim to identify the source and prevalence of such activity within the metaverse space. To do so, we analyzed various dimensions and provided answers to research questions.  In the next section, we will focus on specific dimensions to uncover answers to our research questions in section~\ref{sec:Problem Statement}. Namely, the dimensions we cover with our analysis are (1) communication files and referrer files activities in the metaverse domain, (2) metaverse coins market capitalization, (3) malicious activities in Metaverse coins, and (4) metaverse coins with fiat currency to cryptocurrency.

\section{Results and Findings}\label{sec:results}
Our main results, which analyze and map the relationship between malicious detections in metaverse domains and other artifacts, will be presented in this section. 

\subsection{Communication and Referrer Files in the Metaverse Domain} The popularity of online platforms is determined by the number of visitors, transactions, and overall traffic. Facebook, for instance, boasts billions of registered users and experiences a significant amount of communication and transactions. These interactions are facilitated through manual website exploration, file exchanges, and website database access. However, it is important to exercise caution as autonomous programs such as bots can also interact with these systems. They can inject messages or code, store data in databases, and even remotely manipulate and hijack systems. Therefore, it is crucial to implement proper security measures to prevent unauthorized access and protect sensitive information. In the metaverse, communication files play a significant role. We have collected communication files from all domains and are studying their relationship with malicious activities. Our analysis aims to determine if the number of communication files is linked to malicious detections and identify the types of files responsible for such detections. This information will be crucial in developing preventive policies against malicious threats in the metaverse.

\begin{figure}[htb!]
\centering
\includegraphics[width=0.99\textwidth]{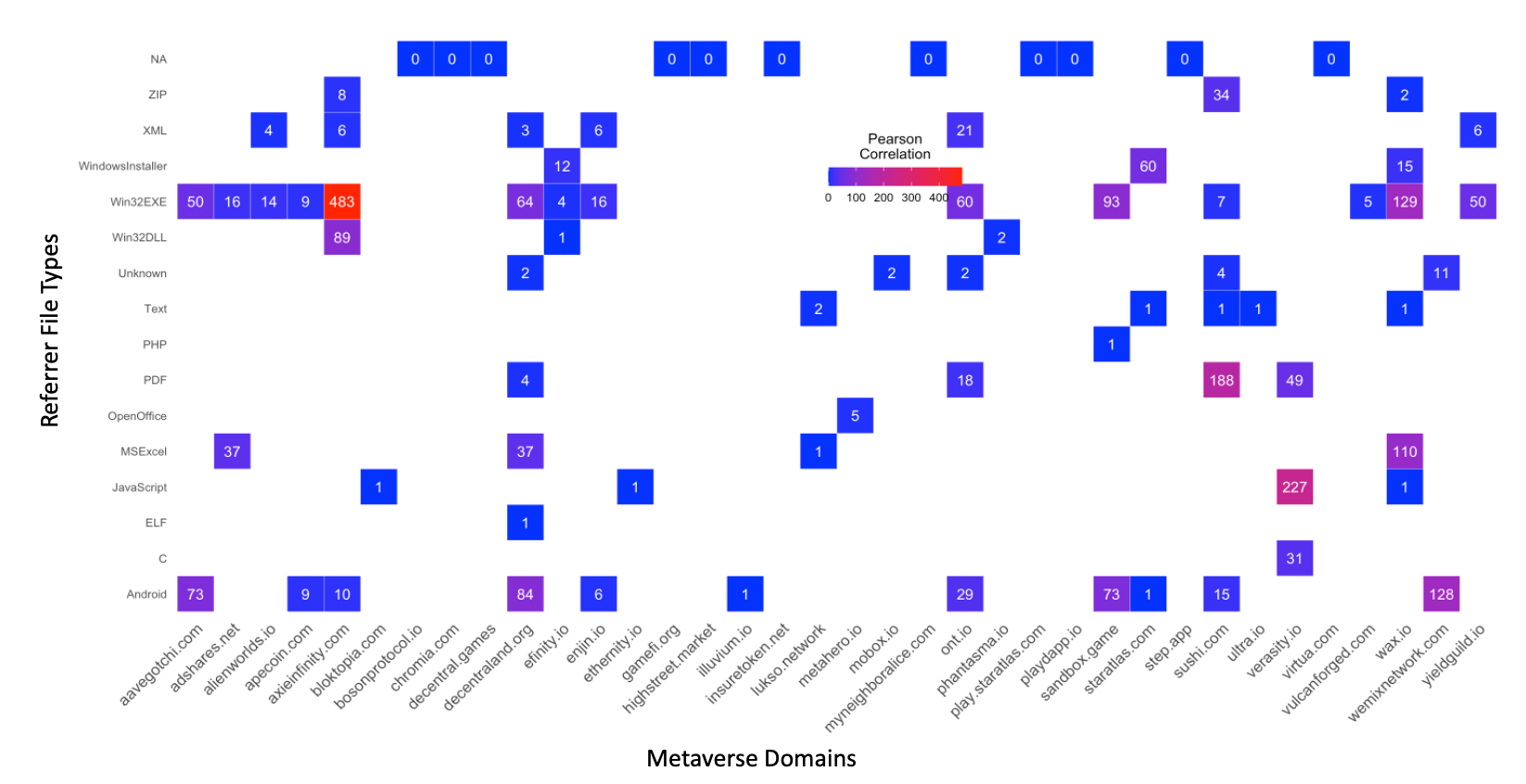}
\caption{Metaverse Domain with Infected Referrer Files}
\label{fig1}
\end{figure}

\begin{figure}[htb!]
\centering
\includegraphics[width=0.85\textwidth]{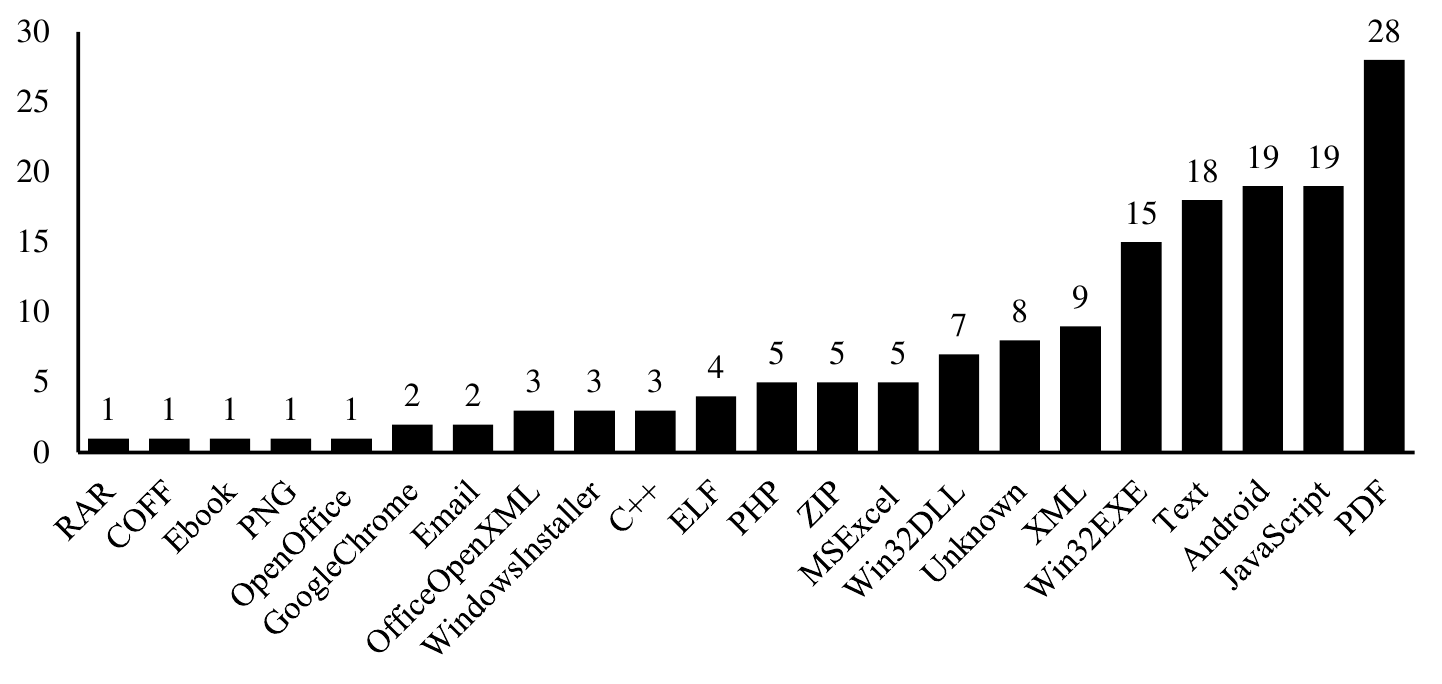}
\caption{Number of Referrer File Types}
\label{fig2}
\end{figure}

\BfPara{Observations} The heatmap in Fig \ref{fig1} displays the frequency of malicious detections in different file types across various domains in the metaverse. The Win32 EXE file type had the highest frequency of malicious detection, with 14 domains recording it. Android came in second, with 11 domains showing a malicious presence. The axieinfinity.com domain had the highest number of malicious detections at 483. Other file types with malicious activity included PDF, Javascript, Android, and MS Excel Spreadsheet. These file types were responsible for most malicious detections in the study.
Additionally, Fig \ref{fig2} shows the frequency of referrer files with no detection. The figure displays a heatmap indicating the frequency of infected referrer file types in the metaverse domain. The number of occurrences for each file type is indicated.

\begin{figure}[h]
\centering
\includegraphics[width=0.99\textwidth]{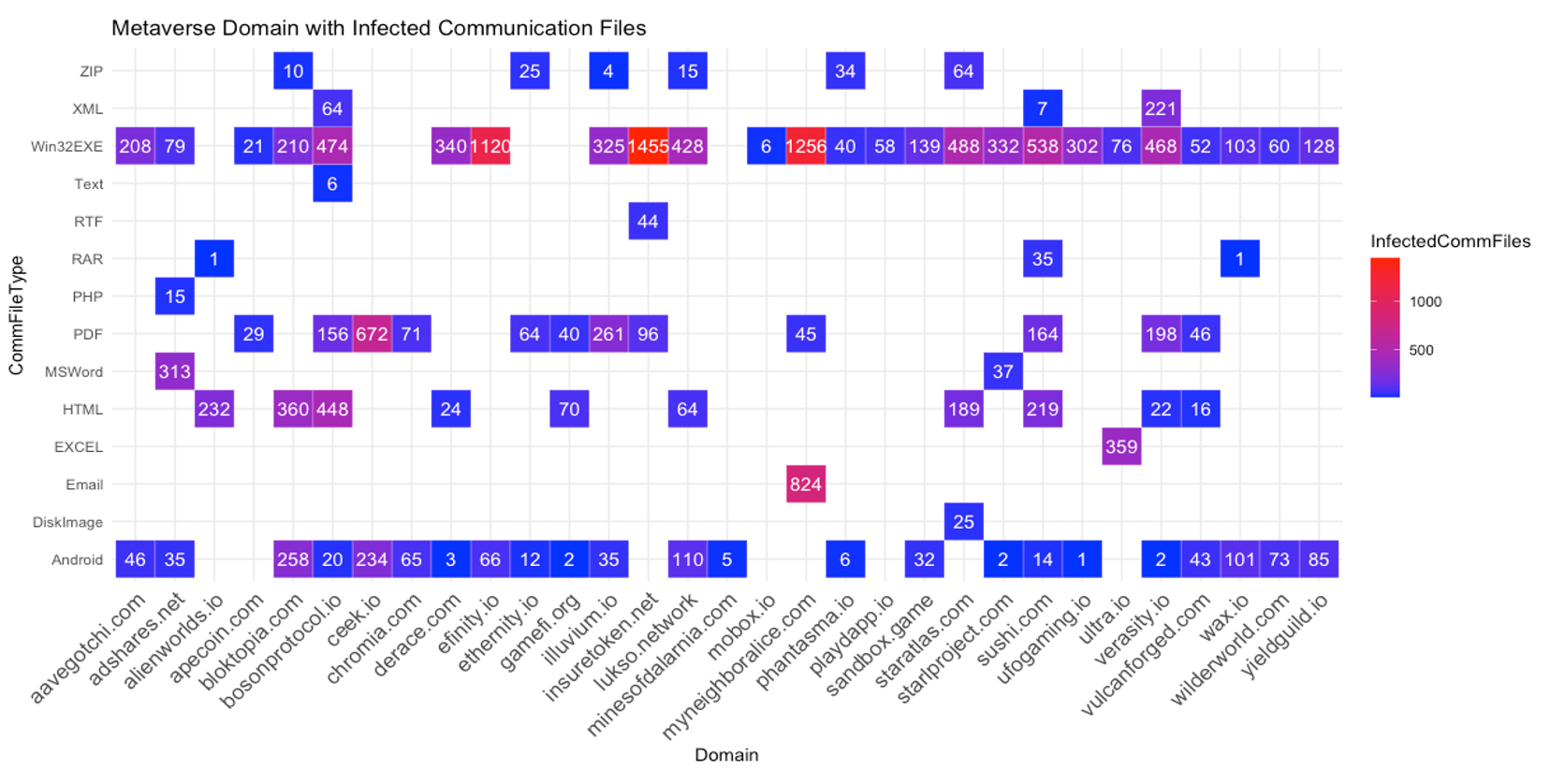}
\caption{Infected Communication File}
\label{fig3}
\end{figure}

The heatmap in Fig \ref{fig3} displays a significant number of communication files with malicious detections. It was discovered that metaverse domains that had malicious detections also had communication files with malicious detections. The Win32 EXE and Android file types were more commonly found than others. The Win32 EXE file type had more detections and was present in approximately 25 out of 31 metaverse domains with malicious detections. Fig \ref{fig3} provides a visualization of the total occurrences of each file type in the metaverse domains, with Android and Win32 EXE file types following the same pattern as previously observed. These two file types are dominant and contribute significantly to the detections recorded in the metaverse domains.

\begin{figure}[htb!]
\centering
\includegraphics[width=.85\textwidth]{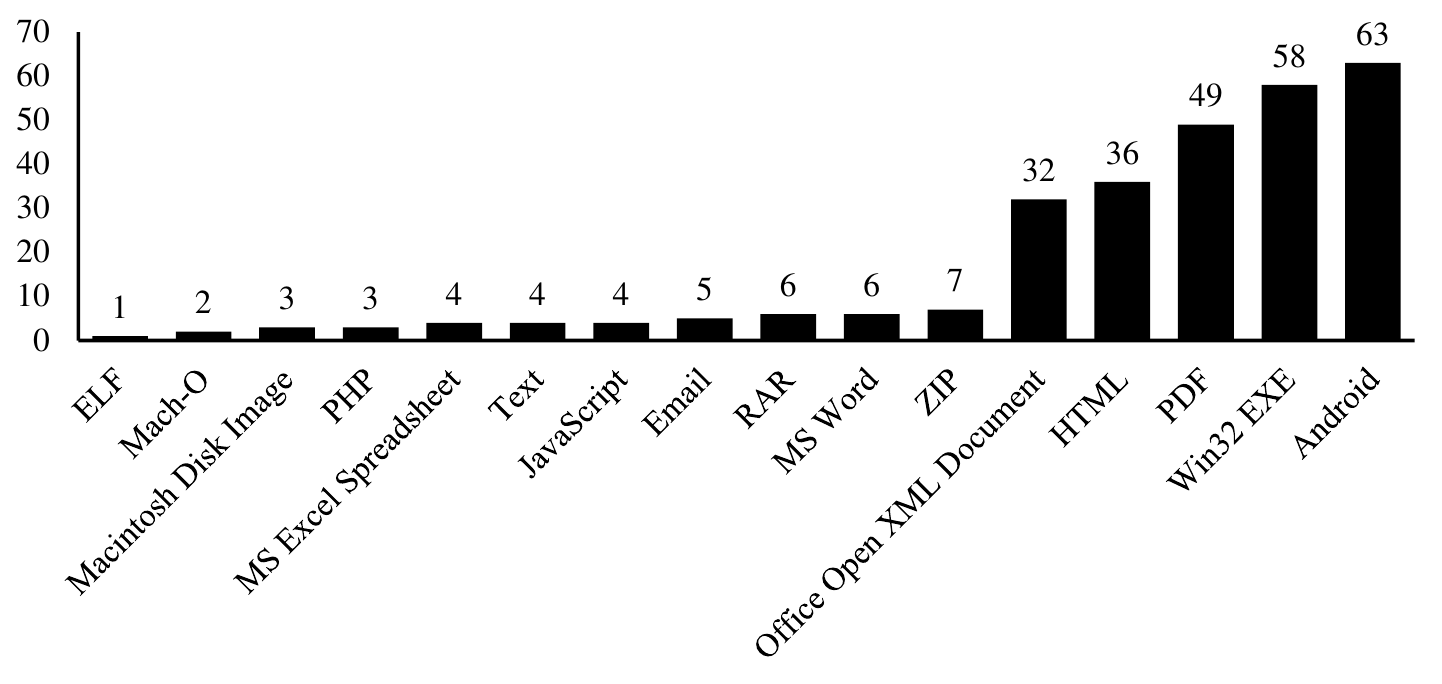}
\caption{Infected Communication File}
\label{fig4}
\end{figure}

\subsection{Metaverse Coins Market Capitalization}  The market capitalization of each metaverse token is obtained from crypto.com\footnote{\url{https://crypto.com/price/categories/metaverse}}. It is important to note that this value is subject to fluctuations, as with other markets. The data provided in this paper reflects the value at a specific point in time and may have since changed. Despite being a futuristic technology, the metaverse already boasts a trillion-dollar market capitalization. The highest-valued token is worth over a billion USD, while the lowest is approximately one thousand USD.

Table 1 shows the list of metaverse tokens in descending order based on market capitalization for the domains with at least 25 million USD capitalization.

\BfPara{Observations} We analyzed the top metaverse token with at least a market capitalization of about 25 million USD for vulnerability and malicious activities by performing a scan with third-party software. The scan result reveals various malicious detections in 31 out of the 44 metaverse domains, representing about 70\% of the domains under consideration as shown in Figure 1 and Figure 3. The malicious detections reported are those obtained from the scan of the metaverse domains, IP addresses, communication files, and referrer files associated with the domains.

\subsection{Malicious Activities in Metaverse Coins} Using Virustotal.com, we conduct thorough scans of files, IP addresses, and domains using many security engines, each utilizing unique algorithms to detect any sign of malicious activity. It is important to note that these engines may classify results differently, which is why we meticulously scrutinize associated components such as passive DNS, communication files, and referrer files to determine the presence of any malicious activity accurately.

\autoref{tab:main}. displays the domains of the metaverse, their corresponding security engines, and the types of malicious detections they can identify. These findings are a result of scanning IP addresses that have been linked to their respective domains.

\begin{table}[htb!]
    \centering\vspace{-7mm}
    \begin{tabular}{|l|l|l|r|}
    \hline
        \textbf{Domain} & \textbf{Security Engines} & \textbf{Type} & \textbf{\# Files} \\ \hline
        playdapp.io & Abusix  & Malicious  & 79 \\ \hline
        playdapp.io & Xcitium Verdict Cloud    & Malicious  & 58 \\ \hline
        playdapp.io & CMC Threat Intelligence  & Malware & 46 \\ \hline
        bloktopia.com & CMC Threat Intelligence  & Malware & 210 \\ \hline
        illuvium.io & CMC Threat Intelligence  & Malware & 261 \\ \hline
        bloktopia.com & CMC Threat Intelligence & Malware & 360 \\ \hline
        step.app & Xcitium Verdict Cloud  & Malware & 544 \\ \hline
        sushi.com & CMC Threat Intelligence & Malware & 588 \\ \hline
        sushi.com & CMC Threat Intelligence & Malware & 655 \\ \hline
        sushi.com & Criminal IP & Malicious  & 124 \\ \hline
        efinity.io & Xcitium Verdict Cloud  & Malware & 680 \\ \hline
        myneighboralice.com & Xcitium Verdict Cloud  & Malware & 822 \\ \hline
        myneighboralice.com & CMC Threat Intelligence           & Malware & 824 \\ \hline
        myneighboralice.com & Xcitium Verdict Cloud  & Phishing & 248 \\ \hline
        myneighboralice.com & Xcitium Verdict Cloud  & Phishing & 840 \\ \hline
        bosonprotocol.io & CMC Threat Intelligence & Malware & 1220 \\ \hline
    \end{tabular}
    \caption{Malicious detection and types}\label{tab:main}
    \label{Malicious types and Security engines}\vspace{-10mm}
\end{table}

\BfPara{Observations} We found eight domains to have malicious infections when the domain IP addresses were scanned. Some domains reported more than one type of malicious detection through different security engines used by virustotal.com. The malicious types in the results are shown in \autoref{tab:main}. Malware, malicious, and phishing are types of files found. The CMC Threat Intelligence security engine was more prevalent, appearing eight times. The table shows the relationship between the metaverse domain and communication files. Every domain that has malicious detection records corresponding communication files. The communications files have shown to have some files with malicious detection, and these files will invariably infect the host domain with malware, phishing, and other maliciousness.

\begin{table}[htb!]
    \centering
    \begin{tabular}{|l|l|l|}
    \hline
        \textbf{Security Engines} & \textbf{Malicious Type} & \textbf{Count of Malware} \\ \hline
        CMC Threat Intelligence  & Malware & 7 \\ \hline
        Xcitium Verdict Cloud & Malware & 3 \\ \hline
        Xcitium Verdict Cloud & Phishing & 2 \\ \hline
        Xcitium Verdict Cloud & Malicious & 1 \\ \hline
        Abusix & Malicious & 1 \\ \hline
        CMC Threat Intelligence  & Malware & 1 \\ \hline
        Total & ~ & 15 \\ \hline
    \end{tabular}
    \caption{Security engines and Malicious types }\label{tab:main}
    \label{Summary of Security engines and malicious types}\vspace{-10mm}
\end{table}

\subsection{Metaverse coins with fiat currency to cryptocurrency} 
Fiat currency in the metaverse refers to using government-issued currencies, such as traditional national currencies (e.g., USD, EUR, JPY) or digital representations of those currencies within virtual worlds or virtual reality environments.

While virtual worlds primarily operate with their virtual currencies or tokens, some platforms or virtual marketplaces may support the integration of fiat currency as a means of exchange. This integration lets users purchase virtual assets or participate in economic activities using real-world currencies. 

Cryptocurrency in the metaverse refers to using digital currencies, typically using blockchain technology, within virtual worlds or immersive virtual environments value~\cite{DBLP:conf/icc/XuLLZS022}. Cryptocurrencies offer a decentralized and secure means of conducting transactions and can play a role in facilitating economic activities within the metaverse.

Categorizing metaverse domains into two groups is crucial for identifying which currency type is more susceptible to malicious activity. These groups include those using fiat currency and those using cryptocurrency. It is important to understand the vulnerabilities associated with each type of currency within these domains.

\BfPara{Observations} After analyzing 44 domains, it was found that 21 of them (48.84\%) use fiat currency. Both classifications of domains showed evidence of malicious activity. It was observed that domains using fiat currency did not exhibit any distinct behavior from those using cryptocurrency, nor did it impact market capitalization. The exchange of fiat currency and cryptocurrency in the metaverse domain is considered a potential factor contributing to malicious activity, but the analysis revealed otherwise. 

\begin{table}[htb!]
    \centering
    \begin{tabular}{|l|l|l|l|}
    \hline
        \textbf{Domain} & \textbf{Fiat Currency} & \textbf{Domain} & \textbf{Fiat Currency} \\ \hline
        apecoin.com & No & minesofdalarnia.com & Yes \\ \hline
        decentraland.org & No & myneighboralice.com & Yes \\ \hline
        axieinfinity.com & No & efinity.io & Yes \\ \hline
        sandbox.game & No & insuretoken.net & Yes \\ \hline
        enjin.io & No & bloktopia.com & Yes \\ \hline
        wemixnetwork.com & No & yieldguild.io & Yes \\ \hline
        sushi.com & No & staratlas.com & Yes \\ \hline
        ont.io & No & virtua.com & Yes \\ \hline
        illuvium.io & No & aavegotchi.com & Yes \\ \hline
        wax.io & No & ufogaming.io & Yes \\ \hline
        lukso.network & No & adshares.net & Yes \\ \hline
        playdapp.io & No & gamefi.org & Yes \\ \hline
        highstreet.market & No & starlproject.com & Yes \\ \hline
        chromia.com & No & play.staratlas.com & Yes \\ \hline
        vulcanforged.com & No & wilderworld.com & Yes \\ \hline
        decentral.games & No & step.app & Yes \\ \hline
        ceek.io & No & ethernity.io & Yes \\ \hline
        mobox.io & No & bosonprotocol.io & Yes \\ \hline
        raca3.com & No & derace.com & Yes \\ \hline
        ultra.io & No & metahero.io & Yes \\ \hline
        verasity.io & No & phantasma.io & Yes \\ \hline
        alienworlds.io & No  &  & \\ \hline
    \end{tabular}
    \caption{Metaverse Fiat to Cryptocurrency}
    \label{Metaverse Fiat to Cryptocurrency}\vspace{-10mm}
\end{table}

\section{Discussion}\label{sec:discussion} Our analysis revealed several instances of malicious activity within metaverse domains. Interestingly, the location of the domains and the DNS and CDN service providers did not contribute to detecting these malicious activities. Our investigation revealed numerous communication and referrer files within the domains, many containing malware. This discovery was unsurprising, as communication and information exchange are common on metaverse web pages. Unfortunately, cyber infections within domains are quite common. Cyber criminals often select their targets based on reconnaissance activities or random selection. With ongoing cyber attacks on cryptocurrency domains and pools, we anticipate similar threats to emerge within metaverse tokens.

\begin{figure}[htb!]
\includegraphics[width=0.95\textwidth]{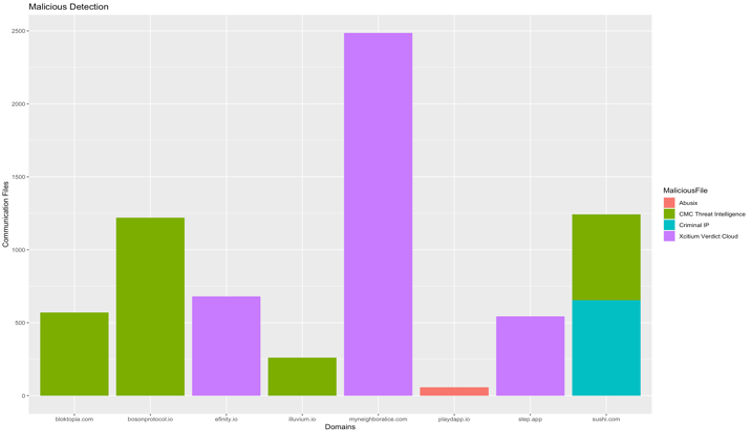}
\caption{Metaverse Domains with Malicious Detection Types}
\label{fig5}
\end{figure}

We have gathered communication files from 44 domains and found malicious activity in 31 of them. However, when we directly scanned the domains and their IP addresses, only 8 out of the 44 domains showed signs of malicious activity, as shown in Fig \ref{fig5}. This means that the number of domains with malicious activity after a direct scan using virustotal.com is much smaller than reported from the communication files and referrer files. It's possible that the large number of communication files with malicious detection does not necessarily translate to domain infections. This could be due to various reasons, such as the domains having security checkpoints, anti-malware, firewalls, or policies that prevent infections from corrupt communication files. While our study doesn't dive deeply into communication files, we can conclude that the eight domains we identified also had communication files with malicious activity.

The website Virustotal.com has its own passive DNS service. We have noticed that the passive DNS results show many malicious detections. Passive DNS stores DNS queries for future analysis, which can help detect malicious networks or infrastructure. However, we cannot confirm if the malicious detections in passive DNS are directly linked to the malicious activities in the eight domains mentioned in Fig \ref{fig5}. It is worth noting that these eight domains are also present in the passive DNS malicious results, as seen in communication files.

\begin{figure}[htb!]
\includegraphics[width=0.90\textwidth]{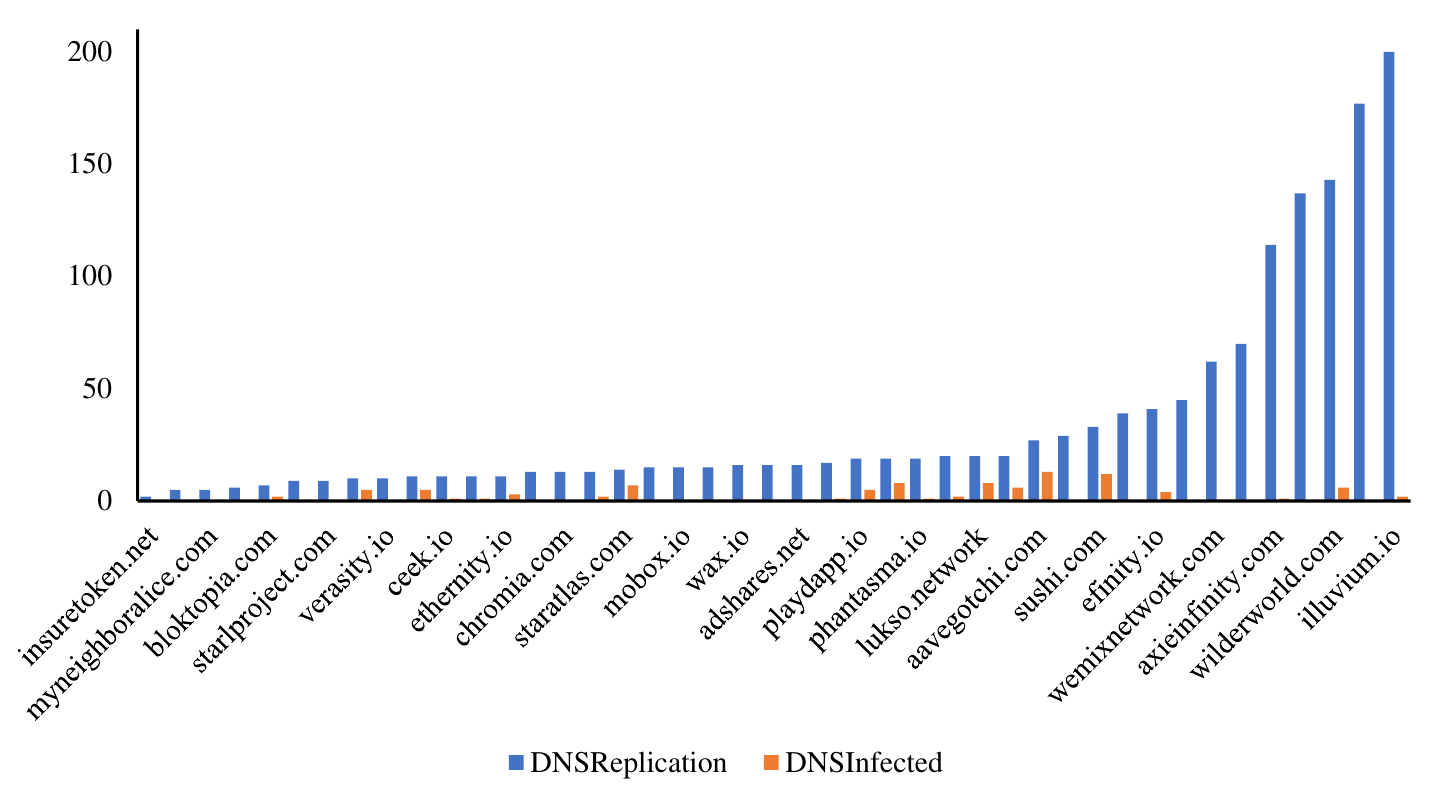}
\caption{Metaverse Domains with Malicious Detection Types}
\label{fig6}
\end{figure}

The body of the analysis is based on several scan results from virutotal.com.

\section{Conclusion and Future Work}\label{sec:conclusion}
Our research analyzes the top metaverse tokens with a market capitalization of at least 25 million USD. We examined the corresponding domains and IP addresses and scanned them for malicious activity using virustotal.com. We found that while many associated files had malicious activity, only 18.6\% of the domains showed signs of maliciousness. Although our analysis confirms the presence of malicious activity in metaverse domains, we were unable to determine the contributing factors. Further research is necessary to identify the sources and factors that contribute to potential malicious activities in the metaverse.

The confirmation of malicious activities in metaverse domains is undeniable, according to the study. It should be noted that a high market capitalization of tokens does not necessarily indicate a lack of maliciousness. The study has identified various forms of maliciousness that must be taken seriously. In the future, we will expand the number and range of metaverse domains for our analysis, expand the study into fiat currencies and association with the security of the metaverse, and further look into the payload (files) in the metaverse platform and their contribution to the security of such systems. 

%\bibliographystyle{splncs}
%\bibliography{reference.bib}

\appendix
\section{Appendix}
Table 1 shows the list of metaverse tokens with at least 25 million USD capitalization.

\begin{table}[htb!]
    \centering
    \begin{tabular}{|l|l|l|l|}
    \hline
        \textbf{Metaverse Token} & \textbf{Metaverse Domain} &\textbf{Metaverse Token} & \textbf{Metaverse Domain} \\ \hline
        Apecoin & apecoin.com & Phantasma SOUL & phantasma.io \\ \hline
        Decentraland MANA & decentraland.org  & Metahero & metahero.io \\ \hline
        Axie Infinity AXS & axieinfinity.com & DeRace DERC & derace.com \\ \hline
        The Sandbox & sandbox.game & Boson Protocol & bosonprotocol.io \\ \hline
        Enjin Coin ENJ & enjin.io & Ethernity Chain ERN & ethernity.io \\ \hline
        WEMIX & wemixnetwork.com & Step App FITFI & step.app \\ \hline
        SushiSwap SUSHI & sushi.com & Wilder World WILD & wilderworld.com \\ \hline
        Ontology ONT & ont.io & Star Atlas & play.staratlas.com \\ \hline
        Illuvium ILV & illuvium.io & Starlink & starlproject.com \\ \hline
        WAXP & wax.io & GameFi GAFI & gamefi.org \\ \hline
        LUKSO LYXe & lukso.network & Adshares & adshares.net \\ \hline
        PlayDapp PLA & playdapp.io & UFO Gaming & ufogaming.io \\ \hline
        Highstreet HIGH & highstreet.market & Aavegotchi GHST & aavegotchi.com \\ \hline
        Chromia CHR & chromia.com & Terra Virtua Kolect TVK & virtua.com \\ \hline
        Vulcan Forged PYR & vulcanforged.com & Star Atlas DAO POLIS & staratlas.com \\ \hline
        Decentral Games DG & decentral.games & Yield Guild Games YGG & yieldguild.io \\ \hline
        CEEK VR & ceek.io & Bloktopia BLOK & bloktopia.com \\ \hline
        MOBOX MBOX & mobox.io & inSure DeFi SURE & insuretoken.net \\ \hline
        Radio Caca RACA & raca3.com & Efinity Token EFI & efinity.io \\ \hline
        Ultra UOS & ultra.io & MyNeighborAlice & myneighboralice.com \\ \hline
        Verasity VRA & verasity.io & Mines of Dalarnia DAR & minesofdalarnia.com \\ \hline
        Alien Worlds TLM & alienworlds.io & & \\ \hline

    \end{tabular}
    \caption{Metaverse Tokens}
    \label{Table 1}
\end{table}

\end{document}